\documentclass[sigconf]{acmart}

\usepackage[english]{babel}
\usepackage{blindtext}
\usepackage{subcaption}
\usepackage{bruce-common}

\newcommand{\ie}{\emph{i.e.}}
\newcommand{\eg}{\emph{e.g.}}

\newcommand*{\Org}{Netflix}
\newcommand*{\org}{Netflix}

\newcommand{\GATE}{\ensuremath{\mathsf{TTE}}}
\newcommand{\TTE}{\ensuremath{\mathsf{TTE}}}

\newcommand{\Reg}{Na\"ive}
\newcommand{\reg}{na\"ive}


\copyrightyear{2021} 
\acmYear{2021} 
\setcopyright{acmlicensed}\acmConference[IMC '21]{ACM Internet Measurement Conference}{November 2--4, 2021}{Virtual Event, USA}
\acmBooktitle{ACM Internet Measurement Conference (IMC '21), November 2--4, 2021, Virtual Event, USA}
\acmPrice{15.00}
\acmDOI{10.1145/3487552.3487851}
\acmISBN{978-1-4503-9129-0/21/11}

\begin{document}
\title{Unbiased Experiments in Congested Networks}

\author{Bruce Spang}
\affiliation{
  \institution{Stanford University}
  \country{USA}
}
\email{bspang@stanford.edu}

\author{Veronica Hannan}
\affiliation{
  \institution{Netflix}
  \country{USA}
}
\email{vhannan@netflix.com}

\author{Shravya Kunamalla}
\affiliation{
  \institution{Netflix}
  \country{USA}
}
\email{skunamalla@netflix.com}

\author{Te-Yuan Huang}
\affiliation{
  \institution{Netflix}
  \country{USA}
}
\email{thuang@netflix.com}

\author{Nick McKeown}
\affiliation{
  \institution{Stanford University}
  \country{USA}
}
\email{nickm@stanford.edu}

\author{Ramesh Johari}
\affiliation{
  \institution{Stanford University}
  \country{USA}
}
\email{rjohari@stanford.edu}
\renewcommand{\shortauthors}{Spang et al.}

\begin{abstract}
When developing a new networking algorithm, it is established practice to run a randomized experiment, or {\em A/B test}, to evaluate its performance. In an A/B test, traffic is randomly allocated between a treatment group, which uses the new algorithm, and a control group, which uses the existing algorithm. However, because networks are congested, both treatment and control traffic compete against each other for resources in a way that biases the outcome of these tests.  This bias can have a surprisingly large effect; for example, in lab A/B tests with two widely used congestion control algorithms, the treatment appeared to deliver 150\% higher throughput when used by a few flows, and 75\% lower throughput when used by most flows---despite the fact that the two algorithms have identical throughput when used by all traffic.

Beyond the lab, we show that A/B tests can also be biased at scale. In an experiment run in cooperation with Netflix, estimates from A/B tests mistake the direction of change of some metrics, miss changes in other metrics, and overestimate the size of effects. We propose alternative experiment designs, previously used in online platforms, to more accurately evaluate new algorithms and allow experimenters to better understand the impact of congestion on their tests.
\end{abstract}

\maketitle

\section{Introduction}

Engineers routinely run A/B tests when testing new network algorithms. In an A/B test, the experimenter randomly allocates a small fraction of traffic (say 1\% or 5\%) to a new algorithm, called the treatment group, and compares its performance against the control group running the old algorithm.
A/B tests are widely used as the gold standard for understanding how a new algorithm will behave at scale. Almost all large tech companies routinely use A/B tests to evaluate changes before deploying them~\cite{KDF+2013,TAOM2010,LRW+2017,CCY+2019,Gov2018,DWM+2019,JC2020a,SM2017,Nit2019}. Networking research often includes the results of A/B tests, and uses them to justify new algorithms~\cite{Iva2020,CK2018,Sha2019,CCY+2019,CCG+2017,Nit2019,KDJ+2020,DRC+2010,FDT+2013,DMCG2011,LRW+2017,JC2020,HJM+2014a,MCD+2020,YAZ+2020,LSBA2021}.

So when we recently ran experiments to test whether \emph{bitrate capping} reduces network congestion for Netflix, we ran A/B tests.
Bitrate capping was introduced in response to COVID-19; major streaming services cooperated with governments to lower bitrates offered and reduce overall internet load~\cite{Gol2020, Ale2020}.
This caused a reduction in congestion in certain networks around the globe. 

We decided to dig deeper, to understand exactly how bitrate capping reduces congestion, and how doing so impacts video quality metrics.
While we had data from just before and after bitrate capping was deployed (and later when it was removed), these were during periods of lockdown and stay-at-home orders when the internet was changing rapidly.
We wanted to conduct a more systematic study of its effects. Naturally, we ran an A/B test where we capped a fraction of traffic to a very congested network. 

In this A/B test, capping didn't appear to reduce congestion at all! In fact, it appeared to make things worse: capped traffic experienced 5\% lower throughput and 5\% higher delay. The A/B test results were so marginal that if we had not had evidence showing that bitrate capping reduced congestion when widely deployed, we might have dismissed it and not explored further. How could a treatment that we \emph{knew} reduced congestion at scale not also reduce congestion in an A/B test? 

Stepping back, we realized the confusion could be caused by \emph{interference}. Interference is when units in the treatment group interact with units in the control group. It is well known in causal inference that interference can bias experiment results \cite{ImbensRubin15}.
In social networks, changing something for a user in the treatment group can impact the behavior of their friends in the control group and bias the results of an experiment~\cite{EKU2016}. In online marketplaces, increasing the price of items in a treatment group can increase the demand for the relatively cheaper items in the control group and bias results~\cite{HLLA2020}. There are many examples of interference bias from markets, education, disease, and more \cite{KR2018, CDG+2013,HS1995,HR2006}. 

Both treatment and control groups in our test used the same network, and their packets traversed the same links and same queues. There is a long line of networking research showing that algorithms compete with each other when sharing a congested network \cite{SJS+2018,Hus2018,WMSS2019,WMSS2019a,CJS+2019,HHG+2018,TKU2019,TKU2019a,BPS+1998,ASA2000,WCL2006,Bri2007a,KRH2020,AWP+2020,DMZ+2018,KJC+2017}. If capping bitrates freed up bandwidth, the uncapped control traffic could take up that bandwidth and get better performance. This could make bitrate capping look worse than it would if the uncapped traffic were not present, even if it was improving congestion. This gave us reason to believe that interference may exist, which would explain our unexpected A/B test results.

In this work we show that interference exists in experiments run in congested networks, and biases the results of A/B tests at scale.
We show that bitrate capping does reduce congestion, and that the misleading A/B test result was due to interference.
In order to do this, we propose and test new experiment designs which more accurately evaluate new algorithms.
Our results suggest that usual A/B testing practice paints an incomplete picture of the performance of new algorithms in congested networks, and should be complemented by additional experiments.

Without interference, A/B tests give us a way to safely and accurately evaluate performance using a very small fraction of traffic.
But because of interference, A/B tests on small fractions of traffic do not accurately predict performance at scale.
Interference therefore creates a tradeoff between safety and accuracy: the only way to accurately measure performance is to run an algorithm on 100\% of traffic, but nobody would do this with an untested algorithm!
Our goal in this paper is to make the networking community, both academic researchers and industry practitioners, aware of this tradeoff and to propose techniques to help mitigate it.
We encourage the community to apply these techniques broadly and evaluate networking algorithms with alternate experiments. We encourage continued measurement and the development of new techniques to mitigate bias.

We begin with an overview of experiment design in Section~\ref{sec:model}. We describe how A/B tests are run, and which quantities they estimate. Using a framework from the field of causal inference, we define the relevant quantities of interest for new networking algorithms.

We then run small lab experiments in Section~\ref{sec:lab} to give examples of how networking A/B tests can be biased. We show that experiments using multiple parallel connections, packet pacing, and different congestion control schemes all exhibit bias. If we were to evaluate these algorithms using \reg{} A/B tests, we would make incorrect conclusions. We might prematurely abandon a good algorithm, or deploy an algorithm that behaves worse when widely deployed than in the experiment.

Returning to our bitrate capping experiments, in Section~\ref{sec:paired} we describe our joint experiments with \org{}. We study the performance of bitrate capping and report on the bias we found in our initial A/B tests. While measurements show that bitrate capping significantly reduces congestion, \reg{} A/B tests do not reflect this behavior. \Reg{} A/B tests miss changes in some metrics, overestimate or underestimate the changes in others, and even get the \emph{direction} of improvement wrong for a few.
We were able to carry out this analysis due to a unique network architecture at Netflix. Using a pair of reliably congested links with well-balanced traffic, we ran different experiments on each link and compared the results.

 Based on our experience, in Section \ref{sec:switchback} we investigate possible ways experimenters can accurately evaluate new algorithms at scale. We discuss two possible paths to managing the tradeoff between safety and bias.  The first is to adapt the common process of gradual deployments to measure interference.   The second involves the use of small-scale, targeted {\em switchback} experiments to more accurately measure the effects of a new algorithm while managing safety concerns. We use the results of our paired link experiment to \emph{simulate} what the experimenter might have obtained in these alternate approaches, and show that both substantially reduce bias.

We believe this paper is just the beginning of work on unbiased network experimentation. There is much to explore in designing more effective experiments, improving the analysis of experiments we run, and understanding the way interference behaves in networks. We wonder how many effective algorithms have been abandoned because of the way we run experiments, and what ineffective algorithms have been deployed because we were misled by A/B tests? Accordingly, we situate our work within the broader context of related research in Section \ref{sec:related} and conclude in Section \ref{sec:conclusion}.
\section{What we want to measure}
\label{sec:overview}
\label{sec:model}

\begin{figure*}
  \centering
  \begin{subfigure}[b]{\columnwidth}
    \centering
    \includegraphics[width=\textwidth]{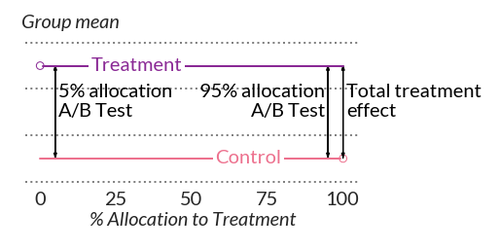}
    \caption{A/B tests without congestion interference}
    \label{fig:effects-sutva}
  \end{subfigure}
  \begin{subfigure}[b]{\columnwidth}
    \centering
    \includegraphics[width=\textwidth]{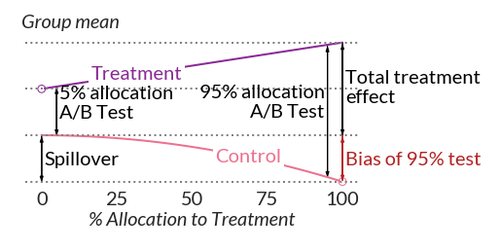}
    \caption{A/B tests with congestion interference}
    \label{fig:effects-no-sutva}
  \end{subfigure}
  \caption{A/B tests are used to estimate the total treatment effect: how much better a treatment is than control if both were deployed globally. A/B Tests give accurate estimates of Total Treatment Effect (\GATE{}) when there is no interference between sessions as in (a), but may be misleading when there is as in (b).}
  \label{fig:effects}
\end{figure*}

Before discussing experiments in more detail, it will be useful to give some background on how they are run, and what they can measure. In this section we provide a formal statistical foundation for A/B testing. The presentation
is borrowed from causal inference~\cite{ImbensRubin15}. The description is simplified, but gives enough conceptual scaffolding for the remainder of our work.

{\bf Treatment assignment.} When we evaluate a new algorithm there are some \emph{units} which run the algorithm. Units may be users, sessions, flows, connections, servers, etc...  We let $U$ be the set of all units. Each unit $i \in U$ is allocated to either {\em treatment} where it runs the new algorithm or {\em control} where it does not. Let $A$ be the vector of treatment assignments to all units. We denote treatment as $A_i = 1$, and the set of treated units as $T$. We denote control as $A_i = 0$ and the set of control units as $C$.

{\bf Potential outcomes.} When evaluating a new algorithm, we are interested in how it improves various metrics. In the language of causal inference, these metrics are called outcomes. Let $Y_i(A)$ be the outcome of interest on unit $i$ given the vector of treatment assignments $A$. $Y_i(A)$ might be the average throughput of unit $i$, the minimum latency, or the 99th percentile packet loss. $Y_i(A)$ can be a random variable, since we expect some variability due to randomness in algorithms and randomness in arrivals.
\footnote{This approach to causal inference via potential outcomes was pioneered by Neyman \cite{splawa1990application} (a 1990 translation of the original 1923 publication) and Rubin \cite{rubin2005causal}; see \cite{ImbensRubin15} for details.}

{\bf Randomized unit assignment.} In an A/B test, we randomly assign units to treatment independently with probability $p$ or control with probability $1-p$. In other words, each $A_i$ is an independent Bernoulli($p$) random variable. We refer to the probability $p$ as the \emph{treatment allocation}.

To make this point more explicit, we introduce some additional notation.  Define $\mu_T(p)$ (resp., $\mu_C(p)$) to be the average outcome value over the randomness in the assignment of treatment (resp. control), when the treatment allocation is $p$:
\[ \mu_T(p) = \E_{T \subset U} \left[ \frac{\sum_{i \in T} Y_i(A)}{|T|} \right]. \]
Depending on the setting and the treatment, $\mu_T(p)$ may or may not depend on the treatment allocation $p$. This is visually depicted in Figure~\ref{fig:effects}. $\mu_T(p)$ is the purple treatment line, and $\mu_C(p)$ is the pink control line.

{\bf Average treatment effect.}  An A/B test evaluates the average treatment effect. This is how much better the treatment group performs than the control group, when a $p$ fraction of the traffic is allocated to treatment and $1-p$ to control. It is defined as:
\begin{equation}
    \label{eq:DIM}
 \tau(p) = \mu_T(p) - \mu_C(p), 
\end{equation}
This is visually depicted in Figure~\ref{fig:effects}. The treatment effect at any point on the graph is the difference between the treatment and control lines.

{\bf Total Treatment Effect.} When evaluating a new algorithm, we are often interested in what \emph{would} happen if we were to deploy it widely.  This is the \emph{Total Treatment Effect}, or $\GATE$: the difference between the average outcome when {\em all} flows are in treatment and when all flows are in control.  In terms of our notation above:
\[ \GATE = \mu_T(1) - \mu_C(0). \]
This is depicted in Figure~\ref{fig:effects}: it is the difference between the right-hand side of the treatment line (when all traffic is treated), and the left-hand side of the control line (when all traffic is allocated to control). Depending on the setting, it may or may not equal the average treatment effect.

Note that this definition of $\GATE$ is from the perspective of the experimenter, and not the internet. The experimenter may only control a small fraction of all traffic on the internet, and in this case $\GATE$ measures what happens if they switched all traffic under their control to a new algorithm. The \GATE{} is also sometimes called the ``global average treatment effect'' in causal inference work (e.g., \cite{KSB+2020}), but we have avoided this name to avoid confusion around this point.

It is also reasonable to talk about \GATE{} in specific groups of traffic. For instance, we may be interested in the \GATE{} if we were to move all traffic globally to a new algorithm, but we may be also interested in the \GATE{} for a single network or a group of networks. This can be incorporated into the definition by changing the set of treatment and control flows.

{\bf Spillover.}  In addition to how well a new algorithm performs on its own, we are often also interested in how a new algorithm impacts existing algorithms. Recently, \cite{WMSS2019} defined the notion of the ``harm'' of a new algorithm, which is the negative effect caused by a new algorithm competing with an existing algorithm. This networking concept is similar to the concept of \emph{spillovers} in the causal inference literature (\eg{} \cite{GXBH2015,CDG+2013}).  Formally, we define the spillover of treatment on control as the effect of increasing the treatment fraction to $p$ on \emph{control} units, relative to when the treatment units were not present. In terms of our notation:
\[ s(p) = \mu_C(p) - \mu_C(0). \]
Spillover is non-zero when deploying a treatment algorithm has some impact on the control algorithm. This is shown in Figure~\ref{fig:effects-no-sutva}. Note that spillover is only defined for $p < 1$. If $p=1$, there is no control traffic and no spillover can occur.

Spillovers may or may not be undesirable. It is possible that deploying a new algorithm can improve existing traffic, and we will see examples of this later.

{\bf Estimating from A/B tests}
All the quantities above are expectations over the distribution of all possible treatment assignments. Any experiment has only one set of treatment assignments and can only observe one set of potential outcomes---all other potential outcomes are missing. The fundamental problem in causal inference is to reason about these missing outcomes given what we observe. 

In causal inference, we use the observed outcomes to estimate the quantities above. An estimator is called {\em unbiased} for some quantity if its expectation is equal to that quantity. 

In an A/B test we randomly allocate units to treatment or control, and measure
\[\widehat{\mu_T}(p) = \frac{\sum_{i \in T} Y_i(A)}{|T|}. \]
This process gives an unbiased estimator of $\mu_T(p)$, since $\E \widehat{\mu_T}(p) = \mu_T(p)$, and similarly for $\mu_C(p)$. By linearity of expectation,
\[\widehat{\tau}(p) = \widehat{\mu_T}(p) - \widehat{\mu_C}(p) \]
is an unbiased estimator for $\tau(p)$, and we can define similar estimators $\widehat{\GATE}$, and $\widehat{s}(p)$.

{\bf Congestion Interference}
In virtually all real-world experiments in networking today, experimenters run an A/B test. They infer that an improvement in the A/B test implies an improvement if the treatment were to be deployed. In our notation, this means that they use $\widehat{\mu_T}(p)$ and $\widehat{\mu_C}(p)$ as an unbiased estimate of the average treatment effect $\tau(p)$, and then interpret $\tau(p)$ as if it were the $\GATE$. This is what we refer to as ``\reg{}" A/B testing.

This process gives an unbiased estimate of $\GATE$ only in the very special case when the outcome of a unit does not depend on the fraction of other units allocated to treatment. This is part of the Stable Unit Treatment Value Assumption (SUTVA) \citep{ImbensRubin15}, and requires that $\GATE = \tau(p)$ for all $p$, and that spillovers are zero for all $p$. Visually, this process assumes that algorithms behave like Figure~\ref{fig:effects-sutva} and not Figure~\ref{fig:effects-no-sutva}.

Any A/B test that runs over a congested network has a clear pathway for interference between units in the treatment and control groups.
Any explicit or implicit change in how the treatment group uses the congested network can create a different network condition for the control groups, which may lead to different behavior. This is \emph{especially} true if the test explicitly changes the timing of how traffic is sent, or the amount of traffic that uses the network. Because of this, we will refer to violations of SUTVA as \emph{congestion interference}.

{\bf Note on averages }
Average treatment effects, spillovers, and \GATE{} are all defined as averages. Average here refers to the distribution of units in the A/B test, and not the outcome metric. The average treatment effect could measure the average difference in average latency, but it could also measure the variance of average latency or 99th percentile latency. Practitioners may also be interested in \emph{quantile} treatment effects, e.g. the difference in 99th percentile latency between treatment and control. These are regularly estimated from A/B test results \cite{Tin2018,AAI2002}. It is straightforward to adapt our definitions to measure quantile treatment effects, and could be done by replacing $\mu_T(p)$ and $\mu_C(p)$ with quantile estimators.

\begin{figure*}
    \centering
    \begin{subfigure}[t]{\columnwidth}
        \includegraphics[width=\textwidth]{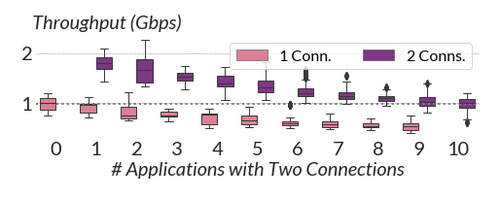}
        \includegraphics[width=\textwidth]{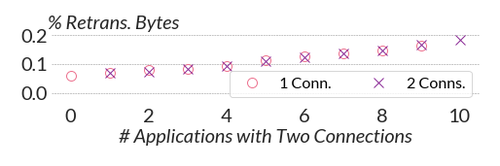}
        \caption{Units are applications using 1 or 2 long-lived TCP connections.}
        \label{fig:parallel-conns-lab}
    \end{subfigure}
    \begin{subfigure}[t]{\columnwidth}
        \includegraphics[width=\textwidth]{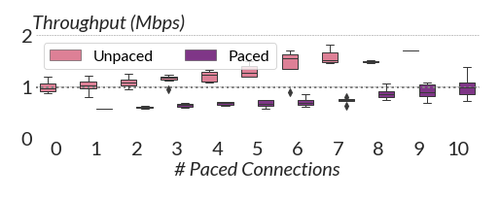}
        \includegraphics[width=\textwidth]{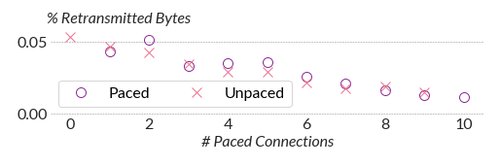}
        \caption{Units are TCP connections which either pace traffic or not.}
        \label{fig:pacing-lab}
    \end{subfigure}
    \caption{Throughput and retransmits in experiments where 10 units share a 10 Gb/s link. Every point on the x-axis is a different A/B test. All tests suggest a large change in throughput and no change in retransmissions, but the difference between 10 treated and 10 control units ($\GATE$) is zero for throughput and large for retransmissions.}
\end{figure*}

\section{Small Lab Experiments}
\label{sec:lab}

When interference is present, \reg{} A/B tests do not accurately describe the behavior of a new algorithm. They mispredict the \GATE{} and give no estimate of spillover. To illustrate this, we set up a small test network in the lab.  The lab setup gives us a global view of how a new algorithm performs at any fraction allocation, and lets us recreate Figure~\ref{fig:effects} for actual algorithms. With these results, we can look at the results of different A/B tests, estimate $\GATE$, and measure spillover. These experiments do not tell us how different algorithms would behave at scale, but they provide easy-to-understand examples of how congestion interference causes bias in \reg{} A/B tests.

\textbf{Lab Setup}
Our lab consists of two servers running Linux 5.5.0, each with an Intel 82599ES 10Gb/s NIC. Each NIC is connected to a port of a 6.5Tb/s Barefoot Tofino switch via $4\times10$Gb/s breakout cables. The switch has a 1 BDP buffer. The sender server is connected to the Tofino with two 10G cables. The interfaces are bonded and packets are equally split between them, which ensures that congestion happens at the switch (otherwise we only see congestion at the sender NIC). We set MTUs to 9000 bytes so the servers can sustain a 10Gb/s rate. We add 1ms of delay at the sender using Linux's traffic controller \texttt{tc}, and use iperf3 to generate TCP traffic.

\subsection{Test 1: Multiple connections}
\label{sec:multiple-connections-lab}

Web browsers, video streaming clients, and other applications request data over multiple TCP connections in parallel. Making simultaneous requests reduces head-of-line blocking, reduces page load time, and increases utilization~\cite{Say2008,Sou2008,Gri2013,GS2019}. This behavior depends heavily on the particular ways an application uses TCP connections and the particular networks it traverses, and so would typically be evaluated with a large-scale A/B test.

However, using multiple TCP connections can also allow an application to outcompete its peers and achieve higher throughput, and so is often called ``unfair'' in the academic literature~\cite{BPS+1998,Bri2007a}. This makes it an ideal example to illustrate how congestion interference can bias A/B tests.

We ran an experiment in the lab to illustrate this behavior and understand the bias it causes. We ran eleven tests in which ten applications used either one or two TCP Reno connections to transfer bulk data. We measured the average long-term throughput and retransmission rates experienced by each application.

Figure~\ref{fig:parallel-conns-lab} shows the results of the lab tests. Each test has two boxplots showing the average throughput for applications using one or two connections. Applications using two connections had 100\% higher throughput and identical retransmission rates than applications using one. As more applications used two connections, their average throughput decreased. When all applications used two connections, their average throughput was identical to when all applications used one. Even worse, retransmission rates were higher when all applications used two connections.

These results are because of the way TCP fairly shares throughput between connections. If $n$ identical TCP connections share a bottleneck link of capacity $C$, we expect each to receive a long-term average throughput of $C/n$. A group of flows with two connections should get a throughput of $2C/n$, 100\% larger than $C/n$. But fundamentally, increasing the number of connections does not increase the capacity of the link so there can be no overall improvement.

This behavior is a well-understood consequence of TCP Reno's throughput fairness. But suppose we followed common practice~\cite{Iva2020,CK2018,Sha2019,CCY+2019,CCG+2017,Nit2019,KDJ+2020,DRC+2010,FDT+2013,DMCG2011,LRW+2017,JC2020,HJM+2014a,MCD+2020,YAZ+2020,LSBA2021} and ran an A/B test to measure how using two parallel connections performed. To illustrate the potential for bias, we will use the same data set interpreted in a different way.

In a na\"ive A/B test, we would randomly allocate some fraction of traffic to treatment and the rest to control. Treatment would use two connections and the rest would use one. We would compare the throughput and retransmissions of the treatment and control groups. \emph{No matter what allocation we picked}, we would see that two connections have a 100\% higher throughput than one, and that there was no impact on retransmission rates. The na\"ive interpretation is that we should always use two connections in production.

\GATE{} and spillover give us a better idea of how two connections perform. The \GATE{} shows that there would be no improvement in throughput and a 200\% increase in the percentage of retransmitted bytes if all traffic were switched to two connections. Spillovers allow us to measure the impact of using two connections on other applications. When nine applications use two connections, the spillovers on the one remaining application using one connection are a 25\% decrease in throughput and an almost 175\% increase in retransmissions.

These results demonstrate that any single A/B test would not accurately measure the impact of changing the number of connections. But we should be careful not to extrapolate too much from the lab results. Applications may benefit from being more aggressive, but using multiple connections can also increase utilization. Without more experimentation, either could be a plausible explanation for a measured increase in throughput.  Fundamentally, we believe that the only way to accurately measure the performance of such a policy would be to run an experiment at scale, on real traffic. We will discuss how to run such experiments later in Section~\ref{sec:switchback}.

\subsection{Test 2: Pacing}
\label{sec:pacing}

Pacing is a generic, widely-used mechanism for reducing packet burstiness in a network \cite{ASA2000, Saeed:2017ky,MLD+2015,CCG+2017}.
With pacing, a host adds delay between successive packets so that it sends a smooth, evenly paced stream of data into the network.

The Linux Kernel has supported pacing for TCP since 2013 \cite{Dum2013,Dum2013a}. It adds delay between successive packets to ensure a rate of $2 \times cwnd / RTT$ during slow start and $1.2 \times cwnd / RTT$ during congestion avoidance \cite{Tor}. 

Prior work, using ns-2, has shown that unpaced TCP traffic outcompetes paced traffic in terms of throughput \cite{ASA2000,WCL2006}. They recommend pacing at a rate of $(cwnd + 1)/RTT$, which is implemented by Linux. These fairness concerns suggest that spillover may be nonzero, which implies that there would be congestion interference in an A/B test.

We ran pacing A/B tests in our lab to measure whether this interference still exists and if it would impact the results of an A/B test. Figure~\ref{fig:pacing-lab} shows the results. Paced traffic (the treatment) obtains 50\% lower throughput than unpaced traffic (the control) in any A/B test, regardless of allocation. In each A/B test, we observed essentially no reduction in retransmissions for pacing.

Applying usual A/B testing practice to these results might have led us to decide not to deploy pacing.
However, if we did deploy pacing, we would be pleasantly surprised to see no impact on throughput and a large decrease in retransmissions.
The A/B tests also miss that pacing is good for other traffic: the spillovers from pacing are an increase in throughput and a decrease in retransmissions.

Pacing highlights the importance of estimating \GATE{} when experimenting with networking algorithms. It is not obvious that pacing changes the way connections compete with each other: we expected it would smooth out bursts and cause lower RTT and loss with no impact on throughput. Without careful experiment design, an experimenter could be easily misled into thinking that pacing is not useful, or waste effort chasing a non-existent bug.

\begin{figure}
  \includegraphics[width=\columnwidth]{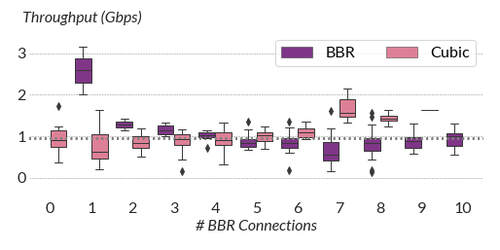}
  \caption{Experiments where 10 TCP connections using Cubic or BBR share a 10 Gb/s link. Throughput is the same if everyone uses either algorithm, but A/B tests suggest that both are improvements.}
  \label{fig:bbr-lab}
\end{figure}

\subsection{Test 3: Congestion Control Algorithms}
There has been extensive study of the fairness of congestion control algorithms (\eg{} \cite{SJS+2018,WMSS2019,WMSS2019a,CJS+2019,HHG+2018,Bri2007a,KRH2020,AWP+2020,Hus2018,DMZ+2018,TKU2019,TKU2019a}). A treatment algorithm is often said to be unfair if it gets a larger share of throughput when competing against a control algorithm. In terms of our metrics, this would be if the spillover on control traffic is a decrease in throughput.

An A/B test will not accurately measure the $\GATE{}$ for an unfair algorithm. The treatment algorithm will take throughput away from the control, making the control perform worse than if the treatment were not present. Most widely-used congestion control algorithms are known to be unfair to at least some other algorithms in certain settings. The resulting biases undermine A/B tests on new congestion control algorithms at scale.

As an example, it's been widely reported that BBR is unfair to Cubic in certain situations \cite{SJS+2018,Hus2018,WMSS2019,WMSS2019a,CJS+2019,HHG+2018}. This unfairness suggests congestion interference, so we ran simulated A/B tests in our lab. We ran ten long-lived TCP connections, and allocated some fraction of them to BBR and the rest to Cubic. Figure~\ref{fig:bbr-lab} shows our results. If we were interested in deploying BBR in this setting and ran a 10\% allocation, we would see a huge improvement in throughput. If instead we were interested in deploying Cubic and ran a 10\% allocation, we would also see a huge improvement! But in this setting there is no difference in throughput between a global allocation to either BBR or Cubic.
\section{Paired link experiment with bitrate capping}
\label{sec:paired}

In response to the increased network usage during the beginning of the COVID-19 pandemic, \org{} worked with various governments to reduce load on the Internet, and rolled out a bitrate capping program which reduced video quality \cite{Flo2020}.  This program capped the video bitrate delivered to clients, while preserving the video resolution based on their subscription plans. It was observed that between March and June 2020, capping the bitrate reduced Netflix traffic in many countries by 25\%, and reduced congestion for a number of ISPs.

In this section, we will describe a controlled experiment we ran to accurately measure the effects of bitrate capping. Given that bitrate capping reduced Netflix traffic by 25\%, we suspected it would decrease congestion. Our preceding lab studies also led us to suspect that standard A/B tests may give biased results. So our goals with this experiment were to:
\begin{enumerate}
    \item Measure the impact of bitrate capping on network performance and video quality of experience, by estimating \GATE{} and spillover effects.
    \item Estimate the bias of \reg{} A/B tests on these measurements, and
    \item Evaluate whether alternate experiment designs would reduce this bias.
\end{enumerate}

These are challenging goals to accomplish simultaneously. To evaluate the bias of a \reg{} A/B test and newer experimental designs, we need to measure what happens when all traffic is treated. But if we treat all traffic, we have nothing to compare against! We could run sequential experiments and compare their results, but this makes strong assumptions about how the system behaves over time. These would be useful assumptions to make when running alternate experiment designs, and we wanted to use this experiment to evaluate these assumptions.

In this section we describe the experiment we ran to achieve these goals. In \org{}'s network, there are a pair of 100 Gb/s peering links to an ISP. The links are reliably congested during peak hours, and are statistically very similar. We treat these two links as ``parallel universes,'' and can compare the outcomes of different experiments to investigate A/B test biases and congestion interference.

Our results are striking and sobering. Bitrate capping reduced congestion at the cost of slightly lower video quality, and improved the performance of uncapped traffic. This was almost completely undetected by \reg{} A/B tests which underestimated some treatment effects, failed to detect others, and, as we will see, even inferred the wrong \emph{direction} of improvement for certain metrics.

\subsection{Paired peering links}
\label{sec:twolinks}

\begin{figure}
  \includegraphics[width=\columnwidth]{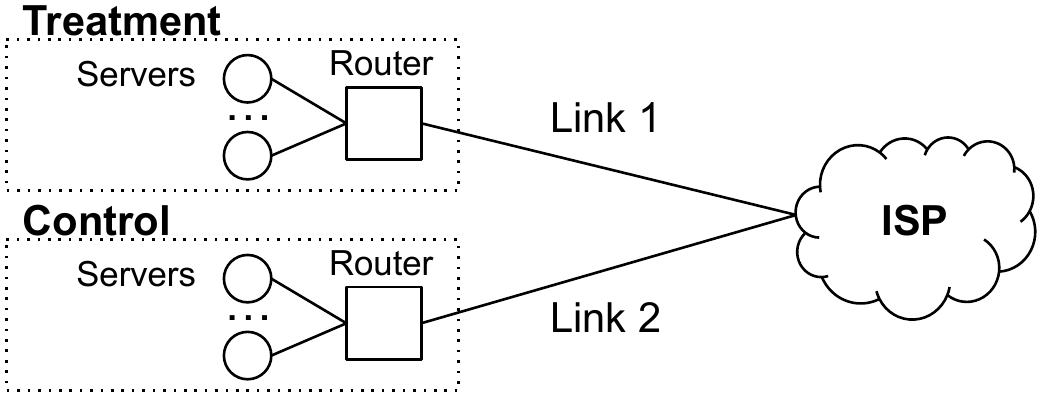}
  \caption{Diagram of the paired link experiment.}
  \label{fig:site-diagram}
\end{figure}

\Org{} has a location with a pair of identical clusters, replicated for scale and redundancy. Each cluster is identically configured with a router and a number of cache servers. Each router connects to a partner ISP via a 100 Gb/s peering link. This setup is depicted in Figure \ref{fig:site-diagram}.

During peak viewing hours, demand from users connecting via this ISP increases until eventually a large standing queue builds up on both links. Latency increases, and throughput and video quality decrease. The congestion has a large impact on the quality observed by traffic, and we suspected strong congestion interference between connections sharing the same link.

{\em A priori}, we are not guaranteed that the two links will be similar to each other, since the system is optimized to serve video and not to run experiments. The content available on the two clusters is not identical, and different traffic is routed to the servers across each link.  To validate statistical similarity between the two links, we collected data on both links during a week-long baseline period, comprising over five million sessions: 50.8\% on link 1, and 49.2\% on link 2. \Org{} collects client- and server-side data on video performance. We looked at 24 important metrics including ones related to network performance (throughput, RTT, etc...) and video QoE (perceptual quality, stability, etc...). For each metric, we used the analysis approach described in Appendix~\ref{app:analysis} to compare links 1 and 2. We will discuss the most relevant subset of these metrics.

We obtained the following results, reported as means and 95\% confidence intervals. Relative to link 2, link 1 had 5\% (0.5\%-10\%) more overall bytes sent, a 2\% (0.1\%-3\%) higher video stability metric, and 0.1\% (0.03\%-0.25\%) lower perceptual quality.  The largest differences were related to rebuffers. Rebuffers are moments when video playback is interrupted because the client is unable to download a piece of video from the server. Relative to link 2, link 1 had 20\% (13-27\%) more sessions with rebuffers; there were four additional metrics related to rebuffers that also exhibited similar differences.  All other metrics did not have statistically significant differences.
Notably, we did not see differences in most metrics we will discuss in our experiment below, including RTT, throughput, video bitrate, cancelled starts, or packet retransmissions.  

Traffic on these links is not perfectly balanced, but it is clearly quite similar. Although the pre-existing differences in rebuffers is large, it is important to note that in absolute terms rebuffers are rare. Given the similarity in other metrics, we believe they are caused by some other difference, such as the content served on the two links. Nevertheless, we carefully discuss our experimental findings regarding rebuffers in Section~\ref{sec:results}, where our observations suggest this difference in fact causes us to \emph{underestimate} the extent to which \reg{} A/B tests are biased.

Being able to run an experiment like this is an extremely unusual situation. Operators work hard to avoid persistent congestion, so it is rare to have a pair of congested peering links. It is even rarer for the traffic to be balanced, and to be able to run separate experiments on each link. \Org{} has hundreds of locations and thousands of peering links worldwide, but only \emph{two} were suitable for this experiment.

\subsection{Experiment design and analysis}
\label{sec:link-experiment}

\begin{figure*}[!ht]
  \centering
  \includegraphics[width=\textwidth]{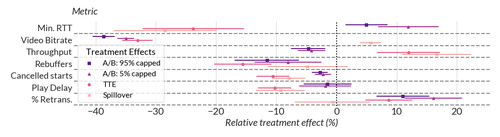}
  \caption{Treatment effects with 95\% confidence intervals in our bitrate capping experiments. Each row is a metric of interest, with the \reg{} A/B Test estimates, and $\GATE$ and spillovers as estimated by the paired link experiment.}
  \label{fig:link-experiment-effects}
\end{figure*}

We now describe the experiment we ran. Our goal was to estimate the effects when most traffic was capped, the \GATE{}, and compare this to the results of A/B tests. We also wanted to measure the spillover of capped traffic on uncapped traffic.

To accomplish this, we ran a pair of A/B tests on the two links. On link 1, we allocated 95\% of flows to treatment ($p = 0.95$). On link 2, we allocated 5\% to treatment.  Computing the \reg{} $\hat{\tau}(p)$ estimator on sessions {\em within} each link allows us to calculate $\hat{\tau}(0.95)$ and $\hat{\tau}(0.05)$.  By comparing the mean of the 95\% {\em treatment sessions on link 1} to the 95\% {\em control sessions on link 2}, we obtain an approximate estimate of $\GATE$. By comparing the mean of the 5\% control sessions on link 1 to the 95\% control sessions on link 2, we can obtain an approximate estimate of the spillover of capping. With this design, we ran A/B tests simultaneously on the pair of links.  The experiment ran for five days, and included about fourteen million video sessions. We analyzed the experiment using techniques described in Appendix~\ref{app:analysis}.

In practice, network experiments are usually run in one of two settings. The first is an initial experiment with a relatively low level of initial treatment allocation, corresponding to the 5\% A/B test. The second is a long-term holdback test, where almost all traffic is treated. We might na\"ively hope that by treating more traffic, we would reduce congestion interference, and this corresponds to the 95\% A/B test.

This experiment may at first appear a bit odd. We are measuring the difference in behavior when \emph{almost} all traffic is capped and \emph{almost} all is uncapped. This is an interesting quantity which tells us a lot about the behavior of bitrate capping during congestion, but it is only an approximation to \GATE{}. The most straightforward way to estimate \GATE{} in this network would be to cap 100\% of sessions on link 1 as treatment, and uncap 100\% of sessions on link 2 as control. We could then compare the means of each group to estimate \GATE{}. However, if we did this, we would have no instances where capped and uncapped traffic shared a link, and we would be unable to compare the results to an A/B test or measure spillover. We could run other experiments other times on the links and compare the results, but we would be making strong assumptions about time invariance. This would require careful experimental design and analysis, and one of our goals here was to \emph{validate} these designs.

Putting it another way: one of our goals is to test the SUTVA assumption, and check whether treatment effects as measured by A/B tests give good predictions of what happens when an algorithm is widely deployed. If SUTVA holds, as in Figure~\ref{fig:effects-sutva}, spillover must be zero, and there must be no difference between the results of the two A/B tests and the approximate \GATE{} we measure. If there is any difference between these quantities in our experiments, SUTVA cannot hold. Knowing that SUTVA does not hold, we would not expect slightly increasing the fraction of capped traffic to fix this problem.

\subsection{Results}
\label{sec:results}

Our results can be summarized as follows: bitrate capping substantially reduced congestion and improved performance of uncapped traffic, and yet the \reg{} estimator would have largely failed to detect this.

Figure~\ref{fig:link-experiment-effects} reports our estimates of treatment effects and 95\% confidence intervals for several important video streaming and network metrics. We report the results of 5\% and 95\% \Reg{} A/B test results (\ie, $\hat{\tau}(0.05)$ and $\hat{\tau}(0.95)$), as well as our estimate of approximate $\GATE$ and our estimate of spillover. The \reg{} estimators are also wrong about the direction of improvement for minimum RTT and average throughput, and the magnitude of average play delay and video bitrate. The spillover is non-zero for most metrics.

Taking the example of average throughput, the two \reg{} A/B tests predicted a 5\% \emph{decrease} in throughput, which na\"ively suggests that capping increased congestion. However, the \GATE{} tells a very different story: that capping \emph{increased} average throughput by 12\%. Spillover shows that capping also benefited other traffic sharing the link: control traffic on the mostly capped link had 16\% higher throughput than that on the mostly uncapped link.

These results can be explained by the way bitrate capping reduced congestion. There was significantly less capped traffic, so it took a larger number of users for the link to become congested. Since user demand was the same on both links, congestion started later, ended earlier, and was less severe on the majority-capped link. The \reg{} estimators were unable to detect this because both capped and uncapped traffic used the same congested link, and therefore saw similar performance.

\begin{figure*}
  \centering
  \begin{subfigure}[b]{\columnwidth}
    \centering
    \includegraphics[width=\textwidth]{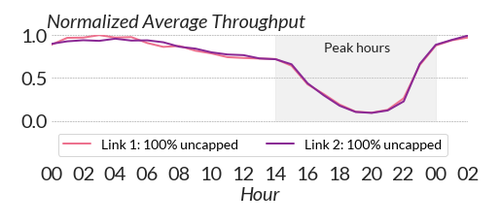}
    \caption{Average throughput for the Saturday of the baseline test period.}
    \label{fig:timeseries-avtp-aa}
  \end{subfigure}
  \begin{subfigure}[b]{\columnwidth}
    \centering
    \includegraphics[width=\textwidth]{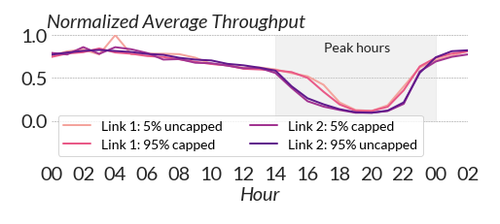}
    \caption{Average throughput for the Saturday of the main experiment.}
    \label{fig:timeseries-avtp}
  \end{subfigure}

  \caption{Client-reported average throughput over time in the experiments, normalized to the largest hourly average. During peak hours, the links become congested and throughput decreases. Capping the majority of traffic in (b) causes Link 1 to be less congested and have higher throughput during most of the peak hours.}
  \label{fig:link-experiment-timeseries}
\end{figure*}

This becomes clearer if we take a closer look at how the average throughput of sessions changes in Figure~\ref{fig:timeseries-avtp}, which can be contrasted with how the behavior during the baseline period in Figure~\ref{fig:timeseries-avtp-aa}. We report the average of all client throughputs during each hour, normalized by the largest hourly throughput. Throughput slowly decreases as overall traffic increases throughout the day, and then suddenly drops when the link becomes congested during peak hours. During the baseline period, there is no difference between throughputs for the two links. During the main experiment, the mostly capped link remains uncongested for longer during peak hours, and has higher throughput before and after the most heavily loaded hours. Despite this difference, the capped and uncapped traffic on the same link have very similar performance.

\begin{figure}
    \centering
    \includegraphics[width=\columnwidth]{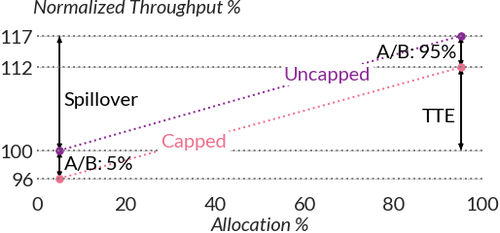}
    \caption{Average values of throughput in the cells in this experiment, with estimands of interest.}
    \label{fig:throughput-effects}
\end{figure}

In Figure~\ref{fig:throughput-effects}, we show the four outcomes of throughput in the experiment: for capped and uncapped traffic as a function of allocation percentage. Both A/B tests confidently report that capped traffic reduces throughput relative to uncapped traffic. However by capping the majority of traffic, we improve throughput for all traffic using the link. This leads to an improvement as measured by \GATE{}, and a positive spillover.

If we considered just one of the A/B tests in isolation, we would falsely conclude that capping traffic makes throughput slightly worse. This is our ``smoking gun''---the confusion arises because treatment and control interfere with each other via congestion on the link.

\begin{figure}
    \centering
    \includegraphics[width=\columnwidth]{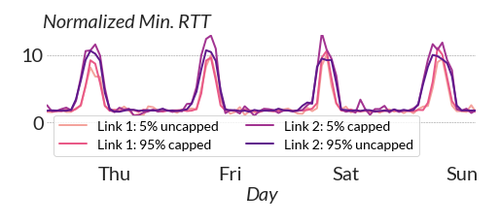}
    \caption{Average of minimum RTT in each connection, normalized to smallest cell value.}
    \label{fig:timeseries-min-rtt}
\end{figure}

We observed similar behavior for round-trip times in the experiment, as shown in Figure~\ref{fig:timeseries-min-rtt}. During congested hours, large queues build up at the congested link, which causes all packets in a session to be delayed, and leads to a sharp increase in the minimum RTT observed during each session. However, because bitrate capping delayed the onset of congestion, the majority-capped link (link 1) had empty queues for more time. The total treatment effect was a 24\% \emph{improvement} in the minimum RTT for the bitrate-capped sessions. The spillover was positive: capping traffic improved the minimum RTT by 27\% for uncapped traffic. Again this was incorrectly estimated by the \reg{} A/B tests which both reported a 5\% and 12\% \emph{increase} in minimum RTT.

We saw similar effects in start play delay, which is the time it takes a video to start playing. This is not surprising: improving throughput and reducing queueing delay should cause videos to load faster. Neither A/B test predicted a significant decrease in start play delay, whereas there was actually a 10\% improvement in total treatment effect. The spillover was also positive: capping traffic reduced play delay by 9\% for both itself and for uncapped traffic.

We measured a 33\% reduction in video bitrate, with positive spillover. Capping the majority of traffic meant that the \emph{uncapped} traffic was able to take up more bandwidth and achieve higher bitrates. It is surprising that despite the spillover, the two A/B tests still give reasonably good estimates of \GATE{}. We believe this is because the majority of the reduction in bitrate comes from the artifical cap, which is applied independently of how other traffic behaves. The spillover is small relative to this effect, but might explain the difference between the 95\% treatment effect and \GATE{}.

\begin{figure}
  \includegraphics[width=\columnwidth]{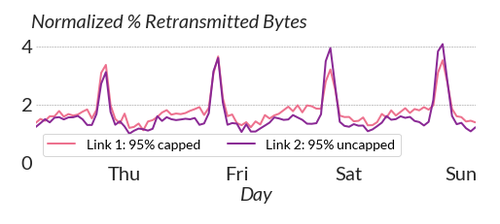}
  \caption{Capping bitrate generally reduced the fraction of retransmitted bytes during congested hours, but caused an increase in uncongested hours.}
  \label{fig:retrans-pct}
\end{figure}

We observed the total treatment effect for capping was a 10\% \emph{increase} in the fraction of sent bytes that were retransmitted. This was driven by a 16\% \emph{increase} in the fraction of retransmitted bytes during off-peak hours, and a 20\% \emph{decrease} during peak hours as shown in Figure~\ref{fig:retrans-pct}. This may seem surprising since bitrate capping reduced congestion, but in fact retransmits did not get worse. Capping reduced the \emph{absolute} number of bytes retransmitted during both during peak and off-peak hours. The apparent increase in the percentage was caused by the absolute number of sent bytes decreasing more than the absolute number of retransmitted bytes. Although odd, \org{} observed similar behavior in a number of ISPs when removing bitrate capping.

Finally, we discuss the impact on rebuffers. Recall from Section~\ref{sec:link-experiment} that we observed a 20\% difference in rebuffers between the links from our baseline analysis prior to the  experiment. Based on our experiment, we believe bitrate capping had at least some impact on rebuffers: we see a 15\% decrease in rebuffers in the A/B tests within each link.  We also measured that rebuffers for the mostly capped traffic in link 1 were 18\% lower compared to the mostly uncapped traffic in link 2.

Given that rebuffer rates were not identical pre-experiment, we investigated further and measured rebuffer rates for both links during the month after we ran the experiment. We consistently found a difference: link 1 had on average 15\% more rebuffers. In 70\% of all hours, and in all but one peak hour, link 1 had more rebuffers than link 2. While we are not certain of the underlying reason for the difference, we believe an 18\% improvement is probably an \emph{underestimate} of the improvement of rebuffers. If we account for the underlying difference between links 1 and 2, it is closer to a 20\%-30\% improvement (rather than 15\% improvement from the \reg{} estimate), suggesting congestion interference.

We conclude by highlighting one reason our results may underestimate the amount of congestion interference. As discussed in Appendix~\ref{app:analysis}, A/B test analysis usually assumes that sessions from different users are statistically independent of each other. By estimating standard errors only on data aggregated to the hourly level, our analysis effectively makes a nearly worst-case assumption that sessions in the same hour are {\em perfectly correlated}. This dramatically increases the size of the confidence intervals we report for \GATE{} and spillover.
\section{Unbiased Experiments at Scale}
\label{sec:switchback}

We care about two different things when evaluating a new algorithm: testing it safely and accurately measuring its performance. We want to experiment safely: if a new algorithm works so poorly that it could cause material harm to the service, we want to detect it quickly and avoid deploying it widely. We also want to be accurate: the goal of a new algorithm is usually to improve some metric, and we need to accurately evaluate whether it succeeded.

A/B tests are used today with the assumption that they are both safe and accurate. If the SUTVA assumption held, we can accurately estimate performance by running an A/B test on a very small fraction of users. This allows us to predict the performance of an algorithm at scale, without broadly deploying a harmful algorithm.

But in the worst case, congestion interference means that an A/B test is neither safe nor accurate. An algorithm which performs well in an A/B test might cause significant harm when it is deployed globally. But if an algorithm has marginal A/B test results and we do not deploy it globally, we may miss out on extremely effective algorithms. 

This is a fundamental tradeoff with congestion interference, and what makes it so difficult to work with in practice. If we want to get a completely unbiased estimate of \TTE{}, we need to allocate 100\% of traffic to a treatment. But for safety reasons we would never allocate 100\% of traffic to an untested or poorly performing algorithm.

In this section, we provide some guidance on how to run experiments in practice. 
We will not be able to completely resolve this tradeoff, but we will describe two ways of measuring congestion interference despite it.

Na\"ive A/B tests are biased in congested networks because of the combination of the A/B experiment design itself, and the flawed causal interference used when interpreting the results of that design. We will propose modifications to the A/B experiment design, and describe the improved causal inference that these modifications allow. First, we propose slightly modifying existing deployment practices to look for congestion interference. This is easy to do and helps build intuition around when congestion interference exists, at the cost of time-related bias and rejecting effective algorithms. To counter this, we also propose running small-scale, targeted switchback experiments to measure how a new algorithm behaves in a specific network.

\subsection{Measure deployed algorithms with event studies}

When deploying an algorithm, it is important to get an accurate estimate of \GATE{}. Optimistically, an algorithm might perform better at scale than it did in small-scale evaluations. Perhaps when an algorithm is run by a larger fraction of traffic, it even further reduces congestion and improves performance than it did in small-scale experiments. Accurately quantifying the improvement is important to understanding its behavior and giving the team working on the algorithm the credit they deserve.

Pessimistically, a new algorithm might perform worse at scale than in small-scale evaluations. This might be a sign of some bug or unexpected behavior in the algorithm, and might suggest it increases congestion or interferes with other traffic on the internet. These are things that are important to know about, so they can be addressed.

Primarily for safety reasons, engineers have developed sophisticated techniques for deploying new algorithms. Engineers gradually deploy changes by slowly increasing the allocation fraction. They continually monitor the system, and stop the deployment if performance degrades.

While engineers typically use gradual deployments to safeguard against failure, they could also be used to conveniently measure the performance of a new algorithm and look for congestion interference. A gradual deployment is effectively a series of A/B tests with treatment allocations ranging from 0\% to 100\%. At each allocation ($p_1$, $p_2$, etc...) we can observe the outcomes for treatment and control. This gives us points on the graph of Figure~\ref{fig:effects}, and we can use these values to estimate the average treatment effect $\tau(p_i)$, the spillover $s(p_i)$, and a \emph{partial} treatment effect $\rho(p_i) = \mu_T(p_i) - \mu_C(0)$. Once the deployment is finished, we can compare 100\% allocation to 0\% allocation and estimate \GATE{}. If there is no interference, for all allocations $i$ and $j$, the average treatment effects are the same $\tau(p_i) = \tau(p_j)$, the partial treatment effects are the same as the average treatment effects $\rho(p_i) = \tau(p_i)$, and there is no spillover $s(p_i) = 0$. We can use statistical tests to check each of these relationships. If they do not hold, it could be a sign of congestion interference.

This is a type of observational design called an \emph{event study} or an interrupted time series \cite[Ch.~11]{KTX2020}. In an event study, we introduce some change, and compare the state of the system before and after. This can be contrasted with a na\"ive A/B test, where we simultaneously compare units with and without the change. In the gradual deployment setting, the change is the increase of treatment allocation from $p_i$ to $p_{i+1}$.

A major flaw with event studies is that it can be difficult to attribute observed behavior to a particular change. This is especially true because of seasonality: holidays, weekends, and political events all tend to have different traffic patterns than other times. Other teams or organizations regularly make changes and deploy software which can affect similar metrics. In the bitrate capping example, we had data from before and after deployment, but chose to run a more controlled experiment to rule out the possibility of other causes for the behavior we observed.

Another flaw is that this process works well for safely deploying new algorithms, but it is heavily biased towards rejecting new algorithms. As an example, suppose we were testing a new algorithm which behaved like the pacing lab experiment in Section~\ref{sec:pacing}. In a small allocation A/B test, this algorithm would look worse: throughput would be down and loss would be unaffected. Seeing this, we might invest our time in other, more promising algorithms. We could slightly increase the size of the allocation to look for interference, but throughput increased quite slowly with allocation size. Even if we were able to detect this interference, it would look small. At this point, we might stop the deployment before the algorithm is able to clearly improve performance.

Despite these flaws, event studies are quick and easy ways to get estimates of \GATE{} and spillovers. Large organizations continually deploy changes. When a deployment happens, it is easy to look at the already-collected metrics and use these metrics to estimate \GATE{} and spillovers. Doing so will help build intuition around which algorithms could be affected by congestion interference.

\subsection{Measure algorithms in development with targeted switchbacks}

Running an event study when deploying a new algorithm is a good way to measure congestion interference and build intuition, but it is a bad way to experiment with new algorithms. We do not want to deploy marginal algorithms to all traffic, and so we may not invest in algorithms that perform poorly in an A/B test. We may miss out on algorithms that have very different effects when widely deployed, like bitrate capping, pacing, or changing the number of TCP connections.

Because of this, we recommend running small targeted experiments in addition to small A/B tests. A targeted experiment allocates a large fraction of traffic within a specific network. The network needs to be structured in such a way that the allocated traffic does not interact with non-allocated traffic. In the paired link experiment in Section~\ref{sec:paired}, we targeted an experiment to two congested links. Using the results from the large fraction allocation, we can get a good estimate of \GATE{} and spillover in this network.

Targeting an experiment allows us to estimate \GATE{} and spillover within a network, without needing to run an algorithm on 100\% of traffic globally. It is standard practice in online platforms \cite{KR2018,SPS+2017}. While we estimate \GATE{} and spillover for a specific network instead of globally, this helps give additional context to A/B test results and improves our understanding of how a new algorithm behaves.

When running these targeted experiments, we recommend using {\em switchback designs}. A switchback design divides time into intervals; a given interval is randomly assigned to be either treatment or control. In a treatment interval, we treat almost all of the traffic with the new algorithm. In a control interval, almost all traffic runs the old algorithm.

At a high level, switchback experiments are analyzed by comparing the treatment and control intervals. While we could do 100\% allocations in these intervals to get a good \GATE{} estimate, we recommend a smaller allocation (e.g. 90-99\%) as in the paired link experiment. Doing so allows us to additionally estimate spillover and the bias of A/B tests, which gives valuable insight into algorithm behavior. The allocation size should be large enough to give statistically significant results, and can be determined by a power calculation.

Like event studies, switchback experiments rely on the change between treatment and control intervals being due to the treatment. However, the assumption is weaker: instead of needing no other events to impact the outcome, a switchback requires that another event does not line up with the treatment intervals.

A switchback experiment can also be vulnerable to carryover effects \cite{glynn2020adaptive,BSZ2021}. The presence of the treatment algorithm can influence the initial conditions of the control algorithm and vice versa. This can cause bias: imagine if we were to switch sessions between one and two parallel connections. Until all sessions that used two parallel connections had completed, the sessions using one would have lower throughput than necessary. If the system reacts poorly to switching between treatment and control, this could also cause problems.

Carryover effects can be mitigated with sufficiently long intervals. However, typically switchback experiments make the worst-case assumption that all sessions in an interval are dependent (see Appendix~\ref{app:analysis} for more details), which essentially means that each interval gives us one data point. Increasing the length of intervals effectively lowers the sample size of the experiment. For networking algorithms, we believe a switch interval of one day is a reasonably conservative place to start. Depending on the setting and the algorithm, it may be appropriate to use a shorter interval on the order of hours or minutes.

\subsection{Evaluating alternate designs}

Our paired link experiment gives us the results of simultaneous, comparable experiments. We previously analyzed that data to estimate \GATE{} and spillovers. We now use it to evaluate event studies and switchback designs, and show that these designs also accurately estimate \GATE{}.

\begin{figure*}[!t]
  \centering
  \includegraphics[width=\textwidth]{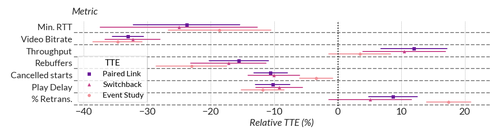}
  \caption{\GATE{} as estimated by the paired link experiment, a switchback experiment, and an event study.}
  \label{fig:effects-by-experiment}
\end{figure*}
\begin{figure}[!t]
  \includegraphics[width=\columnwidth]{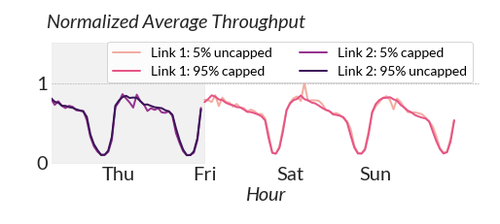}
  \caption{Throughput in a bitrate capping event study. Between Thurs. and Fri., we apply 95\% bitrate capping.}
  \label{fig:event-study-avtp}
\end{figure}
\begin{figure}[!t]
  \includegraphics[width=\columnwidth]{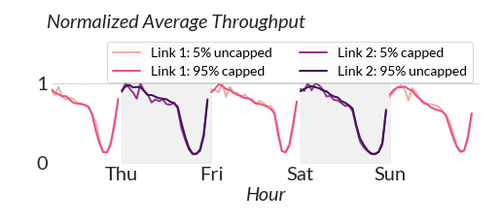}
  \caption{Average throughput over time in a bitrate capping switchback experiment. 95\% of traffic is capped on the first and third and fifth day.}
  \label{fig:switchback-avtp}
\end{figure}

Having two simultaneous experiments allows us to ask what \emph{would} have happened if we ran only one experiment at a time. Our experiment in Section \ref{sec:paired} ran from Wednesday through Sunday, giving us five possible days of data. We can emulate an event study by using data from the 5\% link for a few days and then switching to data from the 95\% link, representing a deployment of bitrate capping to 95\% of traffic. We can emulate a switchback experiment by switching between treatment days and control days more frequently.

We first used baseline data to calibrate a switchback experiment. We ran an A/A test \cite[Ch.~19]{KTX2020} on the paired links in the week following our main experiment: we applied the control to both links and looked for underlying differences. Using the data from the A/A test, we checked that there would have been no false positives with any switchback design. This increases confidence that there isn't a reliable difference between days in a way that would bias the experiment, and we would recommend doing this in most cases.

We also used baseline data to calibrate an event study. We observed that there were false positives in the majority of metrics with any event study in this experiment. We believe this is because weekends tend to have different traffic patterns than weekdays, and an event study must either treat all the weekend days or all the weekdays together. This is an advantage of using a switchback design.

For the event study, we switched to 95\% bitrate capping between Thursday and Friday as shown in Figure~\ref{fig:event-study-avtp}. For the switchback, we alternated between treatment and control, and randomly started with treatment. This assignment is shown in Figure~\ref{fig:switchback-avtp}. All other ways of assigning treatment to days yielded similar results, provided at least one day was in treatment and at least one day was in control.

Figure~\ref{fig:switchback-avtp} shows the average throughput for this example switchback design, which can then be compared with the throughput in the paired link experiment in Figure~\ref{fig:link-experiment-timeseries}. Note that because we are switching between experiments, the clear difference in throughput in the paired link time series is much harder to see in the switchback. This highlights the power of running statistical analyses on switchback data.

Our goal with this approach was to use the clean results from our paired link experiment to demonstrate the power of switchback experiments and event studies. If we had actually run these experiments, the results may have been slightly different. For instance, traffic from both links likely shares some bottlenecks in the provider network during offpeak hours, so it is possible that our results during offpeak hours are biased by congestion interference. However the congestion interference we detect is largely because of the behavior during congested hours on isolated congested links.

\subsection{Results}

The analysis approach for these experiments is identical to the paired link experiment, with the caveat that we only use the subset of the data corresponding to each experiment. We describe the details in Appendix~\ref{app:analysis}.

Figure~\ref{fig:effects-by-experiment} shows the values of \GATE{} estimated by the switchback experiment, event study, and paired link experiment. Both alternate experiments give reasonably good estimates of \GATE{}. The switchback experiment results are very close, and the confidence intervals for its estimates include every \GATE{} from the paired link experiment. It has larger confidence intervals because it includes half as much data. We expect that running the experiment for longer would have reduced the size of the confidence intervals.

The event study gives reasonably accurate estimates of \GATE{} for most metrics, but is biased for throughput, cancelled starts, and \% retransmitted bytes. As we observed in analyzing the baseline data, we believe this is because of seasonality issues: weekends tend to have different behavior than weekdays, and so it is more difficult to attribute the change to the treatment. This is one of the advantages of switchback experiments: randomly choosing intervals over many days helps avoid certain seasonality effects. Despite this, given that event studies are so easy to incorporate into existing workflows, we still recommend cautiously using them to estimate \GATE{} and spillovers when deploying new algorithms.
\section{Related Work}
\label{sec:related}

A/B tests are heavily used in industry research. There recently have been a number of published A/B tests comparing congestion control algorithms, including BBR \cite{Iva2020,CK2018,Sha2019,CCY+2019,CCG+2017}, COPA \cite{Nit2019}, and Swift \cite{KDJ+2020}. There have also been many other published A/B tests for other networking algorithms. These include work on initial congestion windows \cite{DRC+2010}, TCP's loss recovery \cite{FDT+2013}, PRR \cite{DMCG2011}, QUIC \cite{LRW+2017,JC2020}, failure recovery \cite{LSBA2021}, and ABR algorithms \cite{HJM+2014a,MCD+2020,YAZ+2020}. We do not know how congestion interference affected these results.

We are aware of a few published results that include event studies: Dropbox and Verizon both used them to evaluate BBRv1 \cite{Iva2020,Sha2019}, and Google reported one for Timely in \cite{MLD+2015}. In Section~\ref{sec:switchback}, we show how to design and analyze these event studies to measure \GATE{} and spillover, and describe how switchback experiments give more reliable results.

Experiments on router performance, especially those related to buffer sizing \cite{BLG2019,SWH+2019,BGG+2008}, naturally must treat all traffic using the router. Because of this, they tend to have good estimates of total treatment effects.

Recent studies of social network and marketplace platforms have led to improved understanding of causal inference under interference (e.g., \cite{Manski13,Aronow17,Athey18,Basse19,Blake14}), both through novel experimental design (e.g., \cite{Ugander13,johari2020experimental,bajari2019double,BSZ2021,glynn2020adaptive,holtz2020reducing,chamandy16,SPS+2017}) and improved inferential methodology (e.g., \cite{Athey18,Basse16,Basse19,Tchetgen2012OnCI}). We believe our work is the first to show that these issues affect networking experiments and bias their results at scale.

Switchback designs found recent favor as an approach to testing matching and dispatch policies in ridesharing and food delivery platforms, though they have also been used in applications as varied as agriculture \cite{chamandy16,KR2018,BSZ2021,Rob1986,OS1988}. We are unaware of any prior usage of switchbacks in networking.

We have heard some folklore predictions from the networking community that these sort of issues may exist. The only citeable version of this we know of is in \cite{Tow2015}.

Finally, our work is informed by the long line of work on fairness in networking. Unfairness between Cubic and BBR, which we describe in Section~\ref{sec:lab}, was previously reported by \cite{SJS+2018,Hus2018,WMSS2019,WMSS2019a,CJS+2019,HHG+2018,TKU2019,TKU2019a}. Unfairness between parallel connections was first observed by \cite{BPS+1998}. Unfairness between paced and unpaced Reno flows was shown by \cite{ASA2000,WCL2006}. Fairness work is about how algorithms \emph{ought} to share resources, and usually shows that algorithms are unfair in simulations or in a lab \cite{Bri2007a,KRH2020,AWP+2020,SJS+2018,Hus2018,DMZ+2018,WMSS2019a,CJS+2019,HHG+2018,KJC+2017,TKU2019,TKU2019a}. Our work does not address how algorithms \emph{should} share resources, but rather how to avoid experimental bias when they \emph{do}. One way of interpreting our work is as a way to measure unfairness between treatment and control at scale, in production networks.
\section{Conclusion}
\label{sec:conclusion}

Congestion interference biases the results of networking A/B tests at scale, and it is our responsibility as a community to be aware of this phenomenon.  Our results suggest that we should be skeptical when interpreting the results of \reg{} A/B tests, and consider whether alternate experiment designs should be used instead.

As discussed in Section~\ref{sec:switchback}, experimenters can make small changes to existing deployment processes to begin to measure congestion interference, and use targeted switchbacks to further improve these measurements. We should be especially wary of interference when an algorithm changes traffic volumes, tries to control congestion, or is similar to algorithms discussed in the past fairness research in Section~\ref{sec:related}.

We would love to see more work in networking evaluated with congestion interference in mind, either with published switchback experiments, or at least event studies run during a gradual deployment. This is especially true for high consequence proposals, such as new internet standards.

On the research side, there is much more work to be done on evaluating algorithms at scale in congested networks. We encourage further studies to measure bias, in different networks and with different algorithms. We think it would be valuable to design new experiments and analyses specifically for congested networks. The bias of \reg{} A/B tests is both a cautionary tale and a significant opportunity for innovation. The internet surely works better thanks to A/B tests of algorithms run in congested networks. We hope that new algorithms tested with better experiments will help improve it even further.

\section{Acknowledgements}
Thank you to Guillaume Basse and Matthew Pawlicki for the very helpful discussions. Thanks also to Neil Perry, Sundararajan Renganathan, Renata Teixeira, the anonymous reviewers, and our shepherd for all their feedback on the paper.

\def\UrlBreaks{\do\/\do-}
\bibliographystyle{ACM-Reference-Format}
\bibliography{zotero-refs,markets-refs}

\appendix
\section{Ethics}
\label{sec:ethics}
While our experiments involve live traffic running on a large video streaming service, our work is not human subjects research, and we have no way to identify the individual users of the platform. We only have access to performance-related data. We ran experiments which improved behavior during congestion, but they did so at the cost of reducing video quality. \Org{}'s customers have the ability to opt out of experiments, if they choose to.

\section{Appendix: Analysis of experimental data}
\label{app:analysis}

In this appendix we describe our general approach to analysis of data from experiments at scale, and how we apply this approach in the context of the experiments reported in Sections \ref{sec:paired} and \ref{sec:switchback}.  For the duration of the appendix, we consider data for a fixed representative metric collected on a per-session basis (e.g., average throughput).

In our experiments units are video sessions, and we let $A_i$ denote the treatment condition of session $i$, where $A_i = 1$ denotes treatment and $A_i = 0$ denotes control.  Let $Y_i$ denote the observed outcome on session $i$.  Let $h_i \in \{1, \ldots, 24\}$ denote the hour of session $i$.  Our first step in analysis is to aggregate data at the {\em hourly} level: for each hour $t = 1,\ldots,24$ and each treatment condition $A = 0,1$, we compute:

\[ Z_t(A) = \frac{\sum_i Y_i \1_{h_i = t, A_i = A}}{\sum_i \1_{h_i = t, A_i = A}}. \]
This is the average outcome for sessions in treatment condition $A$ during hour $t$.

Next, we use a regression approach to estimate the treatment effect~\cite[Ch~9]{GH2006}, using the following model specification:
$$Z_t(A) = c + \beta_0 A + \beta_t + \eps_i,\ \ \text{for all}\ t, A.$$
Here $t = 1,\ldots,24$ and $A = 0,1$; $\beta_0$ is the coefficient on the treatment indicator; each $\beta_t$ is a fixed effect to control for hour-of-day heterogeneity; $c$ is an intercept term; and $\eps_i$ is the error term.  We fit this model using least squares linear regression, and estimate confidence intervals using Newey-West robust standard errors \cite{NW1987} with a lag of two hours.  This is a common approach in econometrics to account for autocorrelation between successive hours, and heteroskedasticity in the error terms $\eps_i$.  We use hats to denote the corresponding estimates; in particular, $\hat{\beta}_0$ is the estimated coefficient on the treatment indicator, and thus an estimator for the average treatment effect.

\begin{figure}
    \centering
    \includegraphics[width=\columnwidth]{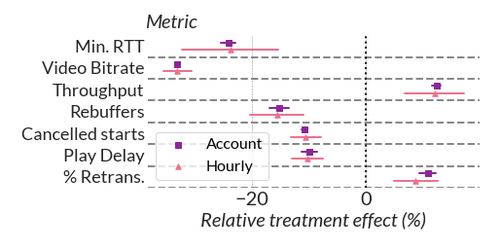}
    \caption{Comparison of treatment effect sizes and confidence intervals when aggregating by hour or by account.}
    \label{fig:gate-by-aggregation}
\end{figure}

We note that the approach we take here---where we aggregate data to the hourly level---essentially makes a worst case assumption that sessions within a given hour and treatment condition are {\em perfectly correlated} with each other.  This is a very conservative assumption, that we feel only strengthens the case in our paper.  Though conservative, this is current practice in analysis of switchback experiments in other industries \cite{KR2018}. If we were to analyze the results using the standard account-level standard errors, we would get much tighter confidence intervals as shown in Figure~\ref{fig:gate-by-aggregation}. Correcting standard error estimates to properly estimate dependencies between sessions remains an active area of investigation.

We now describe how we apply this approach to our experiments in Sections \ref{sec:paired} and \ref{sec:switchback}.

\subsection{Application to paired link experiment}

In Section \ref{sec:paired}, sessions on link 1 were randomized 95\% to treatment and 5\% to control; and sessions on link 2 were randomized 5\% to treatment and 95\% to control.

We carry out four separate analyses on this data.  First, to compute the approximate estimate $\widehat{\GATE}$ for $\GATE$, we consider the 95\% of all sessions in the treatment group on link 1 as our treatment sessions ($A_i = 1$); and the 95\% of all sessions in the control group on link 2 as our control sessions ($A_i = 0$).  We ignore all other sessions.  We then follow the analysis workflow above, and set $\widehat{\GATE} = \hat{\beta}_0$ from the resulting fitted regression.

To estimate spillover, we use only the 5\% control sessions on link 1 and the 95\% control sessions on link 2. We set $A_i = 1$ for the control sessions on link 1, and $A_i = 0$ on link 2. We compute $\widehat{s}(0.95) = \hat{\beta}_0$ from the resulting fitted regression.

Finally we compute two ``na\"ive'' estimates using the difference in means estimator \eqref{eq:DIM} from Section~\ref{sec:model}. In particular, for $p = 0.95$, we use only the sessions on link 1: we consider all sessions in the treatment group on link 1 as our treatment sessions ($A_i = 1$), and all sessions in the control group on link 1 as our control sessions ($A_i = 0$).  All sessions on link 2 are ignored. An analogous approach is carried out for $p = 0.05$ using the treatment and control sessions on link 2 (ignoring all sessions on link 1), to compute $\hat{\tau}(0.05)$. We aggregate to the account level, not the hour level, as is standard when analyzing A/B tests.

Finally, all reported values are normalized to make them more interpretable.  In particular, we divide all estimates by the average across all control sessions on link 2 (where 95\% of the traffic was control).  This approach ensures all reported values are a relative difference measured against the same global control condition.

\subsection{Application to switchback experiments and event studies}

In Section \ref{sec:switchback}, we analyzed a switchback experiment and an event study that was emulated using the data from the paired link experiment.  This analysis was carried out as follows. For the three days chosen to be treatment intervals, we define all treatment sessions on link 1 to have $A_i = 1$, and ignore all other sessions.  For the two days chosen to be control intervals, we define all control sessions on link 2 to have $A_i = 0$, and ignore all other sessions.  We then proceed with the analysis workflow above, and report $\hat{\beta}_0$ as our emulated estimate of $\GATE$.

\end{document}